\documentclass[%
reprint,
 amsmath,amssymb,
 aps,
 pre,
]{revtex4-1}

\usepackage{dcolumn}
\usepackage{bm}
\usepackage{amsmath,amssymb}
\usepackage{graphicx}
\usepackage{color}
\usepackage{csquotes}
\usepackage[percent]{overpic}

\newif\iffigures
\figurestrue 

\newif\ifcomments

\ifcomments
 \newcommand{\KKq}[1]{\textcolor{magenta}{KK:#1}}
 \newcommand{\EFq}[1]{\textcolor{blue}{EEF:#1}}
 \newcommand{\JLBq}[1]{\textcolor{red}{JLB:#1}}
\else
 \newcommand{\KKq}[1]{}
 \newcommand{\EFq}[1]{}
 \newcommand{\JLBq}[1]{}
\fi

\newcommand \df {d_{\mathit{f}}} 
\newcommand \Sc {S_{\mathit{c}}} 
\newcommand \PS {P_{S}} 
\newcommand \Px {P_{x}} 
\newcommand \xmin {x_{\text{min}}} 
\newcommand \xcross {x_{\text{min}}^{cross}} 
\newcommand \PT {P_{T}} 

\begin{document}

\title{Inertia and universality of avalanche statistics: the case  of slowly deformed amorphous solids}

\author{Kamran Karimi}
\affiliation{Universit\'e Grenoble Alpes, LIPHY, F-38000 Grenoble, France}
\affiliation{CNRS, LIPHY, F-38000 Grenoble, France}

\author{Ezequiel E. Ferrero}
\affiliation{Universit\'e Grenoble Alpes, LIPHY, F-38000 Grenoble, France}
\affiliation{CNRS, LIPHY, F-38000 Grenoble, France}

\author{Jean-Louis Barrat}
\affiliation{Universit\'e Grenoble Alpes, LIPHY, F-38000 Grenoble, France}
\affiliation{CNRS, LIPHY, F-38000 Grenoble, France}


\begin{abstract}
By means of a finite elements technique we solve numerically the dynamics of
an amorphous solid under deformation in the quasistatic driving limit.
We study the noise statistics of the stress-strain signal in the steady state plastic
flow, focusing on systems with low internal dissipation.
We analyze the distributions of avalanche sizes and durations and the density of shear
transformations when varying the damping strength.
In contrast to avalanches in the overdamped case, dominated by the yielding point
universal exponents, inertial avalanches are controlled by a non-universal damping
dependent feedback mechanism; eventually turning negligible the role of correlations.
Still, some general properties of avalanches persist and new scaling relations can be proposed.
\end{abstract}

\maketitle

\section{Introduction}
\label{sec:intro}

Avalanche behavior is ubiquitous in nature.
Several systems respond to a slow constant driving with
intermittent dynamics.
Examples are:
Barkhausen noise in ferromagnets~\cite{BarkhausenPZ1917},
stick-slip dynamics in the plasticity of solids~\cite{DasturMTA1981},
earthquakes~\cite{RuinaJGR1983},
creep of magnetic domain walls~\cite{RepainEPL2004},
serrated stress of driven foams~\cite{CantatPF2006} and
crack propagation~\cite{BonamyPRL2008}. 
The common phenomenon is that different regions of a heterogeneous system
are loaded towards instability thresholds; a region that first gets unstable
yields, destabilizes others, and this goes on producing an ``avalanche''.
Recently, such avalanche dynamics has been  evidenced in the time series
of the stress tensor in deformation experiments of
amorphous systems, such as grains, foams or metallic
glasses~\cite{AntonagliaPRL2014,KrisponeitNcomm2014,DenisovNcomm2016}, and has attracted
considerable theoretical interest~\cite{LinPNAS2014,budrikis2015universality,LiuPRL2016}.

Global quantities linked to such collective behavior are
usually power law distributed and allow for the understanding of the system
parameter dependencies in terms of scaling functions.
This is a gift from the self-organized criticality (SOC) paradigm, in which 
stationary states with ``critical'' behavior are
attractors of the dynamics~\cite{BakPRL1987,MaslovPRL1994,PaczuskiPRE1996} and
dominate the scenario as soon as a balance between branching and killing
probabilities is fulfilled.
Among the factors that may break-down this balance in SOC we count
inertia~\cite{JaegerPRL1989,PradoPRA1992,SchwarzPRL2001,KhfifiPRE2008}.

Various works have addressed inertial-like effects in the context of SOC,
like sand-pile models with threshold weakening~\cite{PradoPRA1992,KhfifiPRE2008} 
or depinning-like models that include stress overshoots~\cite{SchwarzPRL2001,MaimonPRL2004}
or softening~\cite{DahmenPRL2009,PapanikolaouPRE2016}.
Moreover, inertia has been explicitly considered on the Burridge-Knopoff
model~\cite{CarlsonPRL1989,CarlsonPRA1989,CarlsonPRA1991}
in the context of seismic faults.
In all cases, avalanche size distributions were found to deviate from the
critical power-law scaling.
It should be said, nevertheless, that none of this models was intended to
illustrate the effect of inertia in the deformation of amorphous solids. 

In this respect, recent studies on atomistic simulations of glasses under
deformation~\cite{SalernoPRL2012,SalernoPRE2013} argue, on the contrary,
that inertial effects drive the system to a ``new underdamped universality class'',
rather than taking it away from criticality.
On the other hand, the same kind of atomistic approach~\cite{NicolasPRL2016},
as well as more coarse grained method~\cite{KarimiPRE2016},
have signaled a strong contrast of the finite-shear-rate rheology between
the overdamped and the underdamped cases; with no signs of universal behavior
in the latter.
Strongly inertial underdamped systems tend to produce, in particular, a
non-monotonic flow curve and the associated localization of the
deformation~\cite{KarimiPRE2016,NicolasPRL2016}.

In this work, we study the influence of inertia in the statistics
of avalanches at the yielding point.
By means of a finite-elements based elasto-plastic model we analyze
the stress time series of a sheared system and relate the stress-drops
there observed with avalanches of several sites yielding in a collective
process.
At high damping we recover the critical avalanche statistics found in 
models of amorphous solids with overdamped dynamics~\cite{TalamaliPRE2011,LinPNAS2014,budrikis2015universality,LiuPRL2016}.
When damping is decreased, we observe that the avalanche statistics
smoothly deviates from the critical distribution.
Even when a power-law shape is observed for a range of avalanche sizes,
the distribution ceases to be scale-free,
developing a bump at large values set by the system size.
Also the exponent of the power-law regime systematically changes as
damping is decreased, indicating a departure from universal behavior.

We understand the avalanche statistics of systems with low dissipation 
as a combination of two distinct kinds of events.
On one hand, avalanches dominated by inertial effects, typically large in size,
frequently system-spanning and with a rather well defined size,
that populates the bump or peak of the distributions.
On the other hand, smaller and more localized
avalanches, which mostly contribute to the power-law regimes
of the distributions, remain controlled by the critical point
attractor of the overdamped dynamics and show little influence
of inertia.

We analyze the emergence of this dichotomy by a careful study of the finite
system-size scaling of the avalanche distributions, their dependence on damping,
and the changes on avalanches geometry.
Within this new scenario, we justify the observed alteration by inertia
of the probability gap for the density of shear transformations 
and propose a scaling relation that links it to the avalanches geometry.

\section{Model and protocol}
\label{sec:model}

\begin{figure}
\begin{center}
\includegraphics[width=0.67\columnwidth, clip]{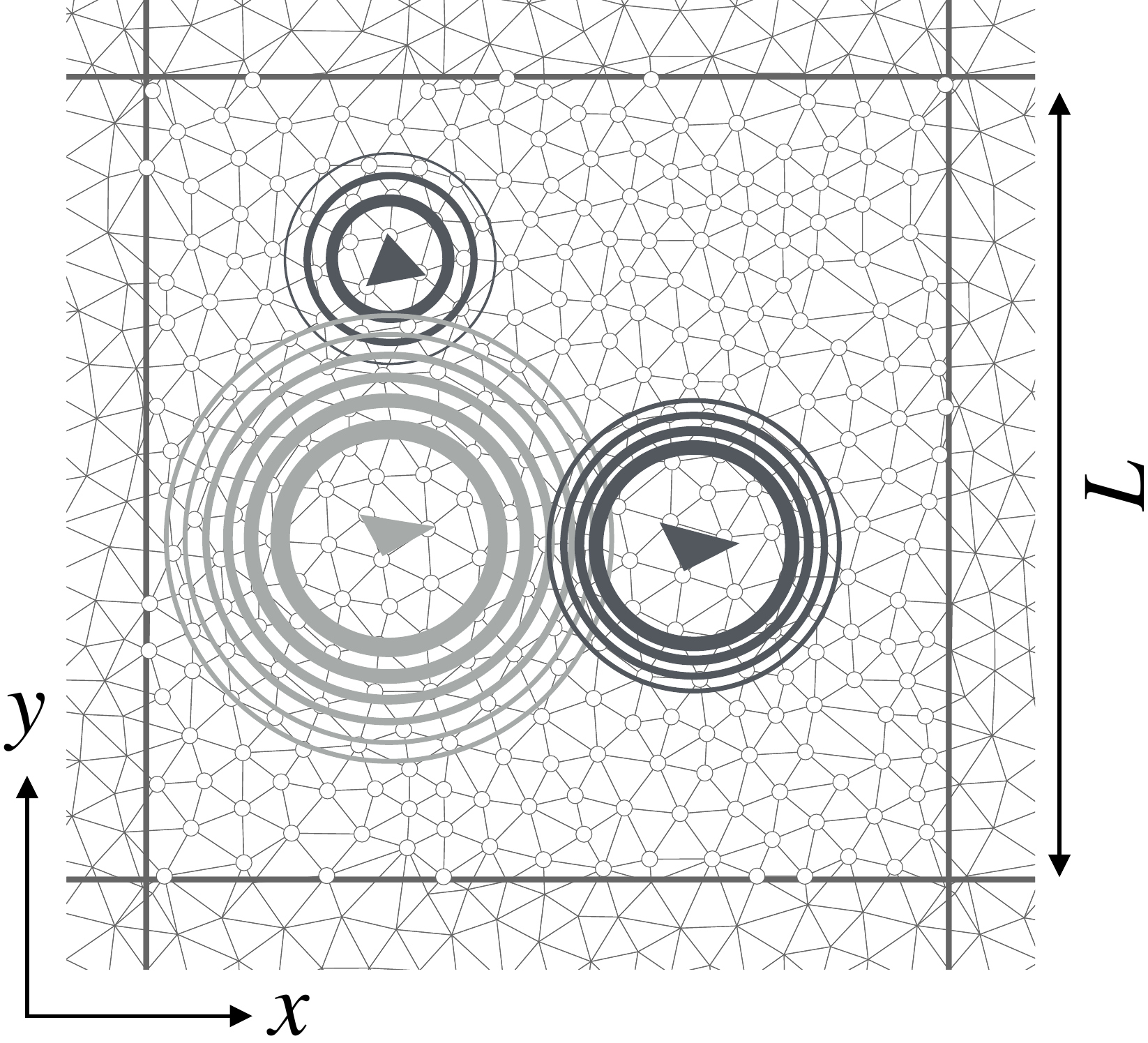}
\caption{\textit{Illustration of the simple-shear set-up:}
Periodicity is imposed along $x$ and $y$. 
A typical size of the finite-elements mesh is $h$ (here $L \approx 20\times h$).
The filled triangle in light gray illustrates an initially yielded element that emits a stress signal represented by annuli of varying thicknesses and diameters. Darker colors denote subsequent yieldings and their corresponding signals.} 
\label{fig:mesh}
\end{center}
\end{figure}

We perform a finite-element-based simulation of amorphous systems in $d=2$ as in \cite{KarimiPRE2016}.
Consider a continuous medium with a displacement field $\bm{u}(\bm{r})$ from an equilibrium
configuration with position $\bm{r}$. 
The equation of motion for the displacement field ${\bm{u}}(\bm{r},t)$ in such a continuum reads
\begin{equation}
\label{eq:wave}
\rho\ddot{\bm{u}}(\bm{r},t)= \bm{\nabla}.\bm{\sigma}(\bm{r},t),
\end{equation}  
where $\rho$ is the \emph{uniform} mass density and $\bm{\sigma}$
is the stress.
Conventionally, the state of a stress tensor is expressed as
$\bm{\sigma} = -p\bm{I}+ \bm{\sigma}_{\text{dev}} $ with
$p=-\text{tr}(\bm{\sigma})/d$ the hydrostatic pressure,
$\bm{\sigma}_{\text{dev}}$ the deviatoric stress, and
$\bm{I}$ the identity tensor.
We further define the \emph{effective} shear stress as
$\sigma = (\frac{1}{2}\bm{\sigma}_{\text{dev}}:\bm{\sigma}_{\text{dev}})^{\frac{1}{2}}$.
The contribution to the stress at $\bm{r}$ is assumed to depend only
on the gradients of $\bm{u}(\bm{r})$ and/or $\dot{\bm{u}}(\bm{r})$,
more accurately on their symmetric parts ${\bm{\epsilon}}$ and/or $\dot{\bm{\epsilon}}$.
Similarly, the deformation rate may be also decomposed as
$\dot{\bm{\epsilon}} = \dot{\epsilon}_{\text{v}} \bm{I}+ \dot{\bm{\epsilon}}_{\text{dev}}$
with $\dot{\epsilon}_{\text{v}}=\text{tr}(\dot{\bm{\epsilon}})$ the volumetric
strain rate and $\dot{\bm{\epsilon}}_{\text{dev}}$ the deviatioric strain rate.
In this case, the effective shear rate can be expressed as
${\dot \epsilon}=(\frac{1}{2}\dot{\bm{\epsilon}}_{\text{dev}}:\dot{\bm{\epsilon}}_{\text{dev}})^{\frac{1}{2}}$.

Each finite-size element represents microscopically rearranging zones
(as in real particulate packings) at a coarse-grained level.
In their rigid phase, these zones are modeled by a Kelvin-type solid,
while past a yielding threshold $\sigma_y(\bm{r},t)$, a simple
Newtonian fluid is employed.
Given this qualitative picture, the constitutive equations are determined by
\begin{eqnarray}
\label{eq:consEq}
\nonumber \bm{\sigma}_{\text{dev}}(\bm{r}, t) &=& \mu {\bm{\epsilon}}_\text{dev}(\bm{r}, t) [1-n(\bm{r}, t)]
+ \eta \dot{\bm{\epsilon}}_{\text{dev}}(\bm{r}, t), \\
p(\bm{r}, t) &=& K\epsilon_\text{v}(\bm{r}, t),
\end{eqnarray}
with $K$ and $\mu$ the bulk and shear moduli, respectively. 
The first terms on the right-hand side of the equations mimic the elastic
contribution, while the second term, only present for the deviatoric piece,
represents viscous dissipation with the viscosity coefficient $\eta$.
Here $n(\bm{r}, t)$ is equal to one during the fluid phase
-- whose life time is limited by a threshold $\tau_p$ --
and zero otherwise.
We set $\tau_p=\tau^\text{min}_v$ with $\tau^\text{min}_v$ defined here later on.
Fixing this time gives a distribution of (plastic) strains in the range $\%0.5-1$.
Realistically, Eshelby zones undergo strong shear deformations, rather than dilation,
to relax the stress, and are, therefore, nearly \emph{incompressible}.
Incompressibility is enforced in the elastic regime by setting $K/\mu \approx 10$.
As in \cite{KarimiPRE2016}, yield stresses are drawn randomly from an exponential
distribution with a mean value $\bar\sigma_y$ and a lower cut-off $\sigma_y^{\text{min}}$,
reminiscent of structural disorder in glassy dynamics.
Also, we dynamically assign a new random threshold to each element after yielding.

Having defined a proper set of constitutive equations, an \emph{irregular} set of
grid-points and linear plane-strain elasto-plastic triangular elements is then
employed in a $L\times L$ periodic cell with typical grid-size of $h$ to
discretize Eq.(\ref{eq:wave}) in space (see Fig.~\ref{fig:mesh}).
Periodic boundaries are implemented by carefully assigning
(based on images at the ``borders'' of the simulation cell)
a lists of neighboring nodes for each node of the irregular
grid, list that remains fix during the simulation. 
 
An area-preserving (simple) shear is implemented to drive the system in a
\emph{quasi-static} manner; that is, an infinitesimal
strain-step\footnote{Given the range of finite system sizes in this study,
we ensure that our chosen strain-step is fine enough to numerically
resolve onset of each avalanche.} $\Delta\gamma\approx10^{-5}$ is initially
applied to the simulation box each time, followed immediately by a stress
\emph{quench} thanks to the dissipative terms present in both Kelvin and
Newtonian descriptions.
The quench runs until the maximum net force on any grid-point is
less than $10^{-6}$ times the average force.
The damping rate for the viscous term is $\tau_d^{-1} = \eta q^2/\rho$, where
the vibrational frequency is $\tau_v^{-1}=c_sq$ with $c_s=\sqrt{\mu/\rho}$
the transverse wave speed.
Here $q$ is the wave number (spatial frequency).
The largest $q$ corresponds to the inverse element size $h^{-1}$,
from which the the shortest vibrational time-scale $\tau^{\text{min}}_v=h/c_s$.
The numerical time integration during quench is performed by means of the
velocity Verlet algorithm with $\Delta t\ll\text{min}(\tau_d,\tau_v)$.
The dimensionless quantity $\Gamma = \tau_d^{-1} / \tau_v^{-1}$, called
damping ratio or damping coefficient hereafter, quantifies the
relative impact of dissipation.
We expect to capture an overdamped dynamics with $\Gamma\gg 1$,
while $\Gamma\le1$ should in principle lead to an underdamped regime.

 \section{Results}
\label{sec:results}

\begin{figure}
\begin{center}
\includegraphics[width=\columnwidth]{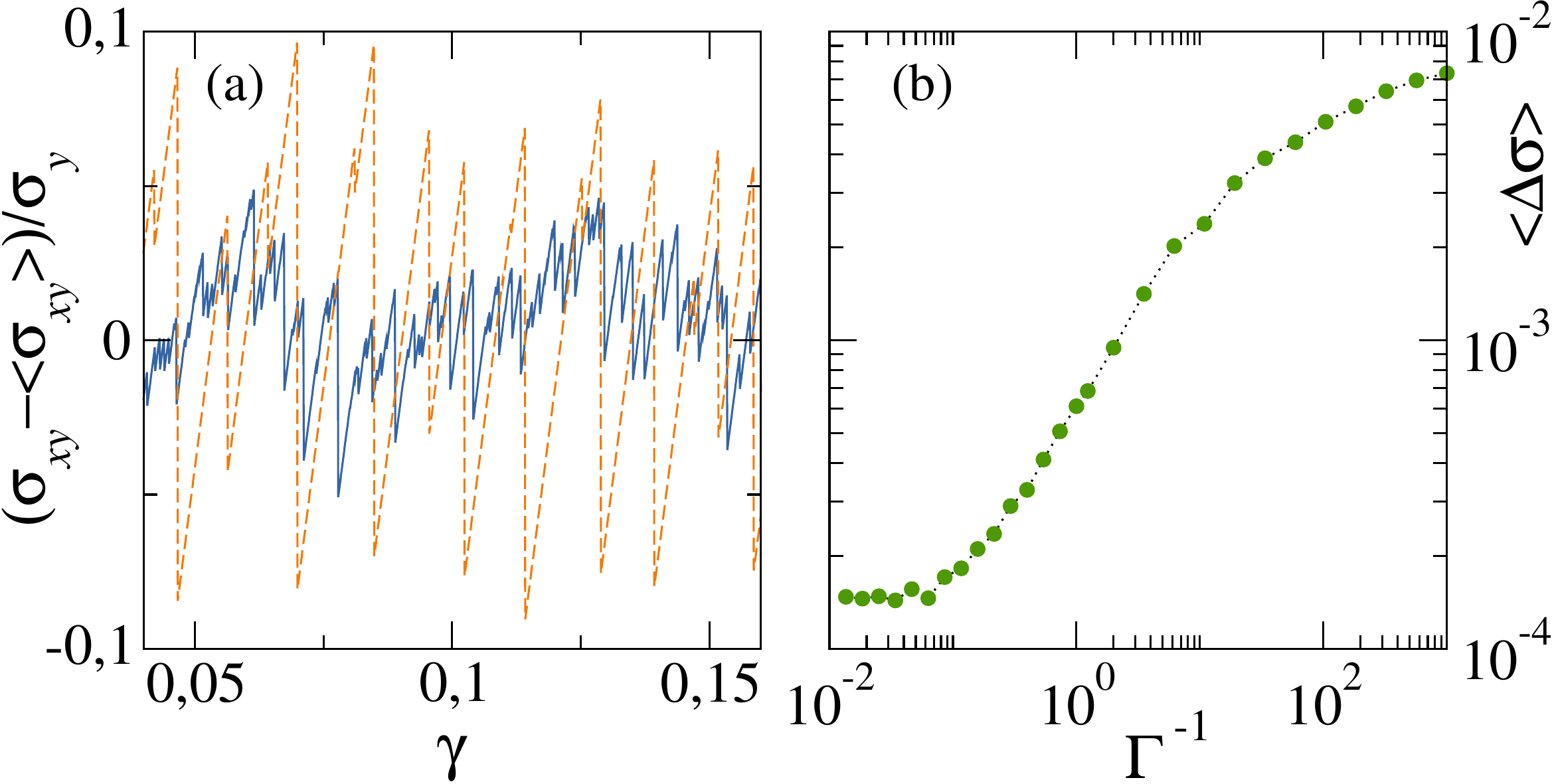}
\end{center}
\caption{\textit{Stress time series and mean-stress drop varying damping}.
(a) Fluctuating stress $(\sigma_{xy}-\langle\sigma_{xy}\rangle)/\sigma_y$
(normalized by mean yield stress) versus shear strain $\gamma$ for two different
damping ratios $\Gamma = 10^{0}$ (blue full line) and $10^{-2}$ (dashed orange line) and
a system linear size $L=40$.
(b) Mean stress drop $\langle\Delta\sigma\rangle$ plotted against inverse damping ratio $\Gamma^{-1}$.
At low $\Gamma^{-1}$ (high damping), the average $\Delta\sigma$ is nearly flat.
It rises only after an inverse critical damping $\Gamma^{-1}_c \simeq 0.2$, that depends
on system size.
}
\label{fig:stressVSgamma}
\end{figure}

As for MD simulations and scalar elasto-plastic models,
imposing a large enough shear deformation results, after a transient regime,  into  a steady-state plastic flow.
The standard observable is the mean stress $\sigma_{xy}(\gamma)$ averaged across the sample.
The stress signal is characterized by a series of elastic loading periods, where the global stress grows
linearly with the applied deformation, interrupted by stress drops $\Delta\sigma$, which are due to plastic
activity and mark an abrupt release of stored elastic energy.
Fig.~\ref{fig:stressVSgamma}(a) illustrates this common scenario in both high and low damping
simulations. A clear qualitative  difference between the high and low damping cases  is evident from the data:
while overdamped fluctuations tend to span a broad range of scales (absence of a characteristic size),
inertial signals seem to be better described by a typical, large, characteristic stress drop,
appearing in a quasi-periodic fashion.
The averaged stress drop value $\left<\Delta\sigma\right>$ as a function of the inverse
damping coefficient $\Gamma^{-1}$ (bigger the more inertial is the system) is shown in
Fig.~\ref{fig:stressVSgamma}(b).
At low values of $\Gamma^{-1}$ the mean stress drop saturates,  which can be used to define   the
overdamped limit.

\subsection{Avalanche size distributions}
\label{sec:avalanche-distributions}

The magnitude of the stress drop reflects the number of sites involved in 
a correlated sequence of yielding events or \textit{avalanche}.
Therefore, in agreement with earlier studies ~\cite{SalernoPRL2012,SalernoPRE2013},
we simply define  an extensive avalanche size as
$S = \left<\sigma_{xy}\right> \Delta\sigma L^d$.
By using the mean stress value $\left<\sigma_{xy}\right>$ as a multiplicative
factor that depends on $\Gamma$, $S$ acquires (within a constant factor)
the dimensions of an energy drop, that better quantifies the collective
behavior and allows for a comparison among systems with different damping.

\iffigures
\begin{figure}[t!]
\begin{center}
\includegraphics[width=0.9\columnwidth, clip]{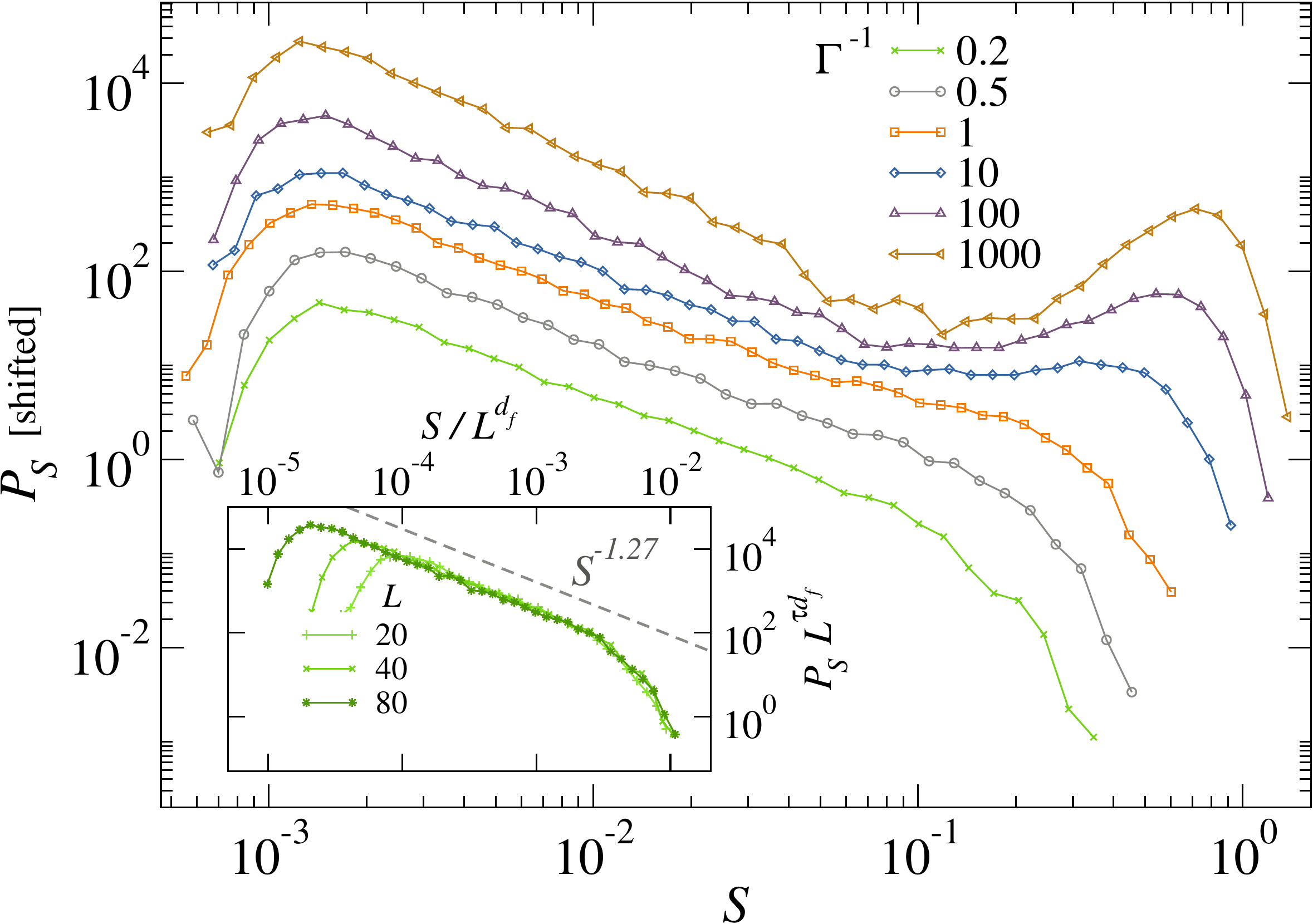}
\end{center}
\caption{\textit{Avalanche size distributions varying damping}.
$\PS$ versus avalanche size $S$ at different inverse dissipation coefficients $\Gamma^{-1}$.
Data correspond to a linear system size $L=40$ (2046 blocks).
Curves for different $\Gamma$ are arbitrarily shifted along the vertical axis for clarity.
A peak develops at large $S$ when inertia becomes important.
Also the exponent $\tau$ in the power-law regime increases.
Inset:
$\PS S^{\tau \df}$ versus $S L^{-\df}$ for the highly overdamped case,
collapsing different system sizes. 
$\tau \approx 1.27$ and $\df \approx 0.95$
} 
\label{fig:PSvaryingDamping}
\end{figure}
\fi

Figure~\ref{fig:PSvaryingDamping} shows distributions of avalanche sizes $S$ for
different inverse damping coefficients 
at a fixed system size $L=40$ built from a statistics of 10000 events.
We first notice that an overdamped system ($\Gamma^{-1}=0.2$) displays,
after a characteristic lower cutoff, 
a power-law decay in $S$ that is later on suppressed by an exponential decay.
In fact,  based on previous results, we expect to observe in this limit
a behavior $\PS \sim S^{-\tau} G(S/\Sc)$ with $G(\cdot)$ a rapidly decaying
function and the cutoff set by $S_c$.
The inset of Fig.~\ref{fig:PSvaryingDamping} shows curves for such a high damping 
at different system sizes rescaled as $\PS S_c^\tau$ vs $S/S_c$, where we introduce
the size dependent cutoff $S_c(L) \propto L^{d_f}$ with $d_f$ the
\textit{fractal dimension} of the avalanches.
A collapse onto a unique master curve is obtained for large $S$. 
The power-law fitting and collapse found are characterized by the values
$\tau=1.27 \pm 0.05$ and $d_f = 0.95 \pm 0.05$, in agreement with
previous results of various elasto-plastic models~\cite{TalamaliPRE2011,LinPNAS2014,LiuPRL2016}
and molecular dynamics simulations in the overdamped limit~\cite{SalernoPRL2012,SalernoPRE2013,LiuPRL2016}.

When the damping is decreased and inertia starts to be relevant,
two main new features become apparent: 
On the  one hand the $\PS$ distributions deviate from a  pure power-law shape;
a shoulder develops at large values of $S$ and evolves into a clear local maximum
at very low damping.
On the other hand, an apparently robust power-law regime survives at smaller
values of $S$; nevertheless the exponent $\tau$ characterizing the decay of $\PS$ systematically
deviates from its overdamped value  as the damping is decreased, reaching values of $\tau \simeq 1.5$ for
the lower damping displayed, with no signs of saturation.
Furthermore, a closer inspection of the  distributions obtained at low damping evidences
the existence of an intermediate regime in which a probability depletion is produced;
simultaneously showing how the inertial peak at large $S$ degrades the free-scale
regime and suggesting a separation between two different kind of events.
Let us recall that, indeed, a differentiation of two groups of events has been
proposed for the Burridge-Knopoff model~\cite{CarlsonPRL1989,CarlsonPRA1989,CarlsonPRA1991}.

\iffigures
\begin{figure}[t!]
\begin{center}
\includegraphics[width=0.9\columnwidth, clip]{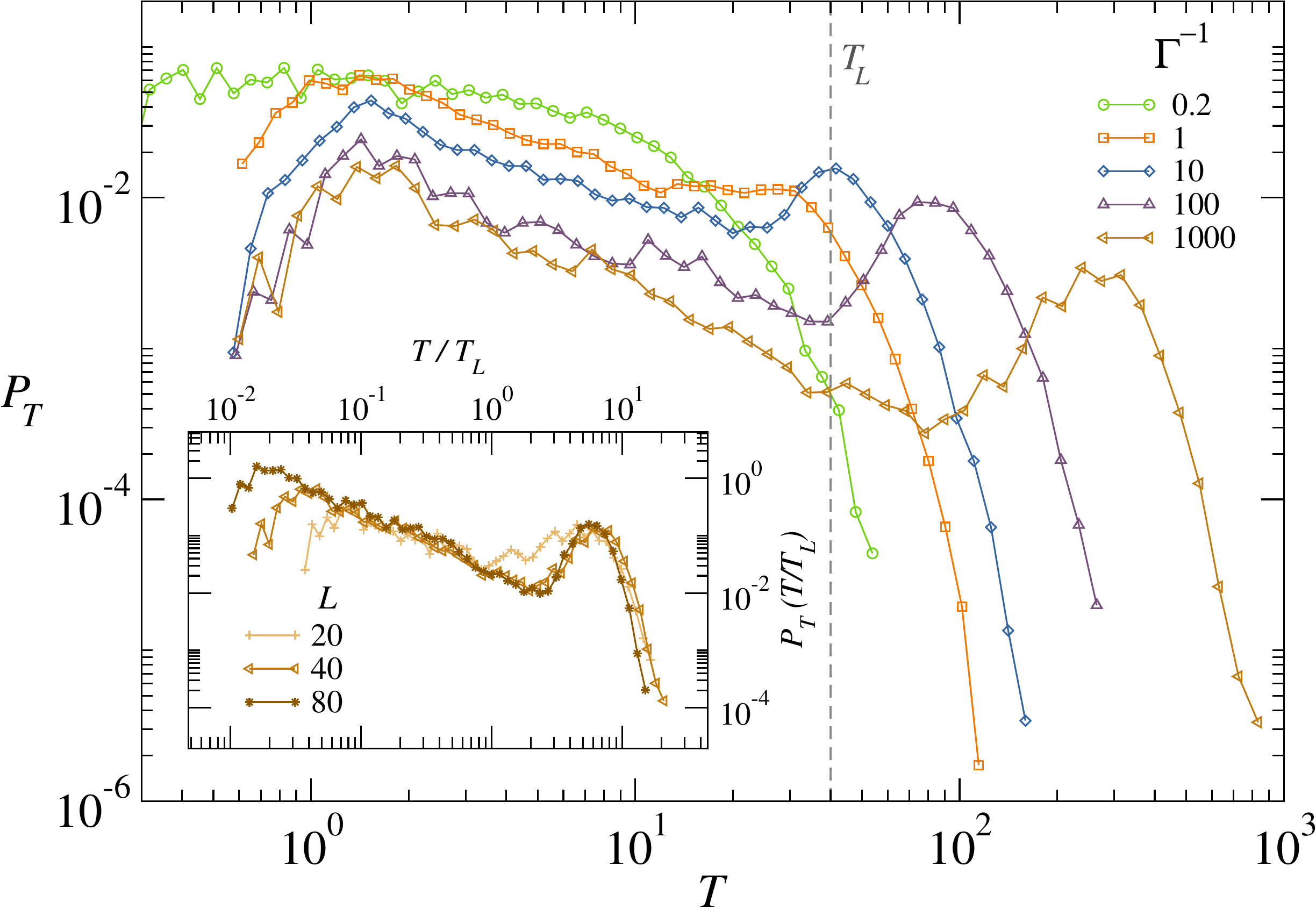}
\end{center}
\caption{\textit{Avalanche duration distributions varying damping}.
$\PT$ versus avalanche duration $T$ normalized by the system-size-dependent
duration $T_L\equiv L/c_s$ at different inverse dissipation coefficients $\Gamma^{-1}$.
Data correspond to a linear system size $L=40$ (2046 blocks).
When inertia becomes important the $\PT$ cutoff overcomes $T/T_L$ and a peak develops.
Inset:
$\PT$ versus $T/T_L$ for $\Gamma^{-1}=1000$ and different system sizes $L=20,40,80$.
} 
\label{fig:PTvaryingDamping}
\end{figure}
\fi

In our case, this distinctive feature of the avalanche size distribution
can be better understood by analyzing the distribution of avalanche durations, $\PT$.
Figure \ref{fig:PTvaryingDamping} shows this distribution 
in units of $\tau_v$, which is the relevant microscopic time for inertial
systems~\cite{NicolasPRL2016};
similar to $\PS$, the distribution shows at large damping 
a power-law regime, exponentially suppressed for long durations.
As damping is decreased, a peak at large values of $T$ appears. 
A vertical dashed line marks the size dependent value $T_L = L/c_s$.

Figure \ref{fig:PTvaryingDamping} suggests a simple interpretation of the
appearance of the inertial peak in the distributions. 
At the local level, the immediate consequence of the lack of dissipation
is a stronger effect of a yield event in its surrounding, that could be
seen as an effective reduction in the local thresholds.
Nevertheless, the most notorious inertial feature, as we understand now,
comes from a long-range action.
When energy is not rapidly dissipated it keeps traveling around the lattice
in the form of shear elastic waves.
Inertial effects become dominant when an avalanche duration is long enough
such that the avalanche is able to reinforce itself through its own elastic
waves (traveling around the system or possibly being reflected  at the boundaries
of a non-periodic system).
We expect this time threshold to be proportional to $T_L$, the time needed
by the elastic waves to propagate across one of the system's main axis.
This is confirmed by the behavior of $\PT$ at different damping coefficients.
In the still overdamped case ($\Gamma^{-1}=0.5$, putting the limit at $\Gamma=1$)),
$\PT$ is already exponentially decaying by $T \simeq T_L$ meaning that a big
majority of avalanches are not affected by this
finite-size dependent inertial effect.
On the other hand, under-damped systems allow for larger durations $T>T_L$,
meaning that a growing number of avalanches are long-lived enough to
sustain their activity by using the energy stored in the elastic waves they emitted.

Although the wave speed does not depend on $\Gamma$, another time scale
relevant for the efficiency of this feedback mechanism is the one governing
the damping of the elastic waves emitted by one event, $T_r$.
This time scale will increase as damping decreases, and also be affected by
the density of active sites that may scatter the wave.
The efficiency of the feedback effect will become stronger and create longer
lasting avalanches as $T_r$ increases compared to $T_L$ and the elastic waves
are able to cross the system repeatedly.
This is illustrated by the drift of the peak in $\PT$ towards larger values
of $T$ as $\Gamma$ decreases.
%

In such a context, it is logic to suspect that the inertial effect observed
may be a simple consequence of a finite system size.
It is appealing to imagine that the characteristic bimodal distribution
of $\PS$ may vanish when increasing the system size and that the
scale-free scaling would persist longer and longer.
Interestingly, this does not happen.
On the contrary, at a fixed low damping and increasing the system size
we observe that the effect of inertia gets more and more marked.

\iffigures
\begin{figure}[t!]
\begin{center}
\includegraphics[width=0.9\columnwidth, clip]{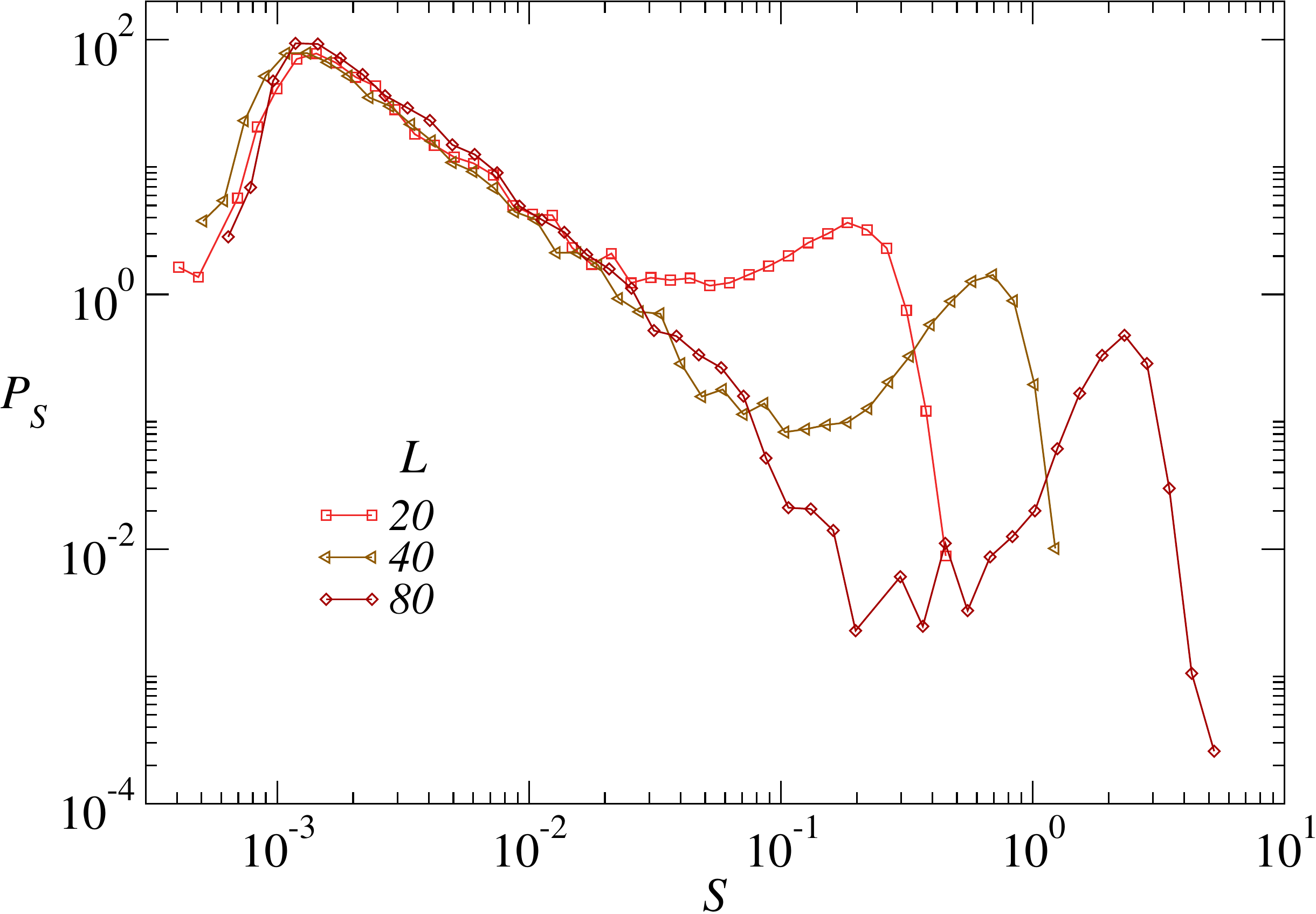}
\end{center}
\caption{\textit{Avalanche size distributions at low damping}.
$\PS$ versus $S$ for $\Gamma^{-1}=1000$ and different system sizes $L=20,40,80$.
Increasing the system size does not diminish the inertia fingerprint in the avalanche
size distribution. 
On the contrary, the characteristic ``bump'' becomes further evident as more massive
avalanches are allowed by a bigger simulation box.
} 
\label{fig:PofSfinitesizeUnderdamped}
\end{figure}
\fi

In Fig.~\ref{fig:PofSfinitesizeUnderdamped} we show the avalanche size
distributions at $\Gamma^{-1}=1000$ and different system sizes.
%
%
As we increase $L$ an occasional persistence of the power-law behavior
is not evidenced.
Instead, a depletion is caused in between the scale-free regime and
the inertial peak and the peak becomes sharper.
Bigger systems, allow for bigger avalanches to develop.
The larger the avalanche, the bigger its mass and increased its chances to be pulled
out from a scale-free size statistics by inertia. 
  
Finally, let us emphasize that the particular role played by the system size
in this discussion is intrinsically related to the use of a quasistatic protocol.
At most one avalanche is taking place in the system at any given instant.
On the other hand, for a system driven at a finite but small strain rate,
the size of the system should be replaced by the strain rate dependent
distance between avalanches. 
 
\subsection{Local distances to threshold}

\iffigures
\begin{figure}[t!]
\begin{center}
\includegraphics[width=0.9\columnwidth, clip]{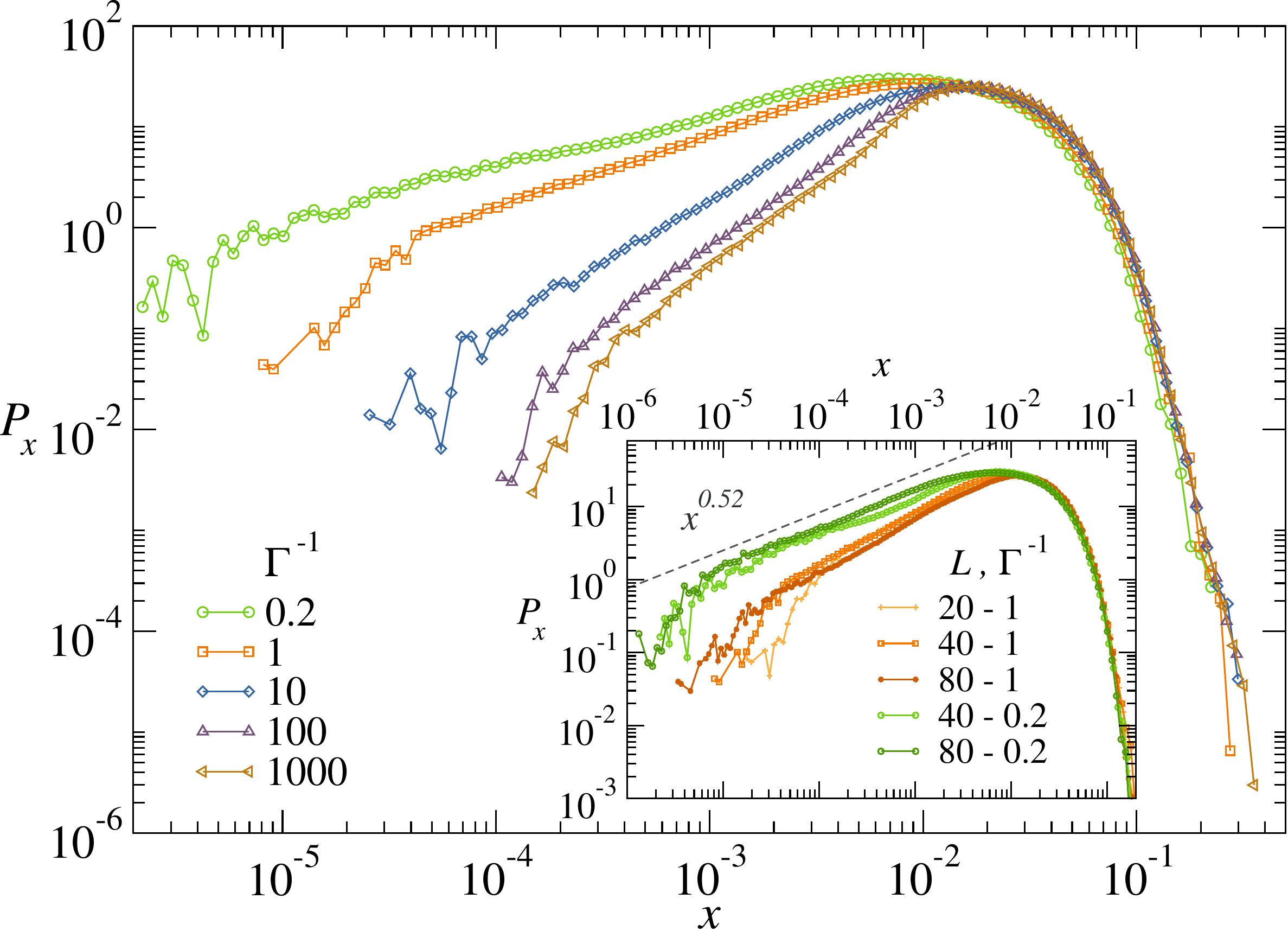}
\end{center}
\caption{\textit{Distances to yielding distributions varying damping}.
$\Px$ versus $x$ at different inverse dissipation coefficients $\Gamma^{-1}$.
Data correspond to a linear system size $L=40$ (2046 blocks).
In the overdamped limit $\Px \sim x^\theta$ with $\theta \approx 0.52$.
As inertia becomes important there is a systematic change in $\Px$.
Inset:
Distributions at different system sizes corresponding to the overdamped case ($\Gamma^{-1}=0.2$)
and a slightly less damped case($\Gamma^{-1}=1$) illustrating the finite size effects.
} 
\label{fig:PXvaryingDamping}
\end{figure}
\fi

A key quantity that characterizes the specific avalanche behavior in plastic solids is 
the distribution  $\Px$ of \emph{local} distances from instability $x_i=\sigma_{y_i}-\sigma_i$,
with $i$ an index running over the $N$ blocks that compose the system.
At each loading stage in the stress time series, the minimum $x$ value in the lattice
denoted by $\xmin$ determines the global instability threshold for the next event.
The extent of an avalanche will be controlled by the full distribution $\Px$
and its evolution  during the stress drop.

It has been shown, following stability arguments~\cite{LinEPL2014}, that the functional
form of $\Px$ near the yielding transition should be a power-law $\Px \sim x^{\theta}$
with $\theta>0$ for small $x$, a \textit{pseudo-gap}.
Moreover, assuming the independence of the $x_i$, a simple scaling argument links the
exponent $\theta$ with the exponents $\tau$ and $\df$ controlling the distribution of
avalanche sizes, $\theta+1=1/\left(1-(2-\tau)\df/d\right)$.
This prediction was found to hold in numerical simulations of overdamped systems,
both in quasi-static protocols~\cite{LinEPL2014,LinPNAS2014} and approaching
the limit of vanishing strain-rate~\cite{LiuPRL2016} from finite values,
giving consistent sets of exponents $\{\theta,\tau,d_f\}$ in two and three
dimensions.
We now report the influence of the damping on $\Px$ by analyzing
configurations right after a stress drop has occurred.

Figure~\ref{fig:PXvaryingDamping} shows $\Px$ for systems with different damping $\Gamma$.
Our estimation of a power law $\Px\sim x^\theta$ in the overdamped case ($\Gamma^{-1}=0.2$)
agrees within error bars with previous estimates of the exponent $\theta$ in 2D
elasto-plastic models~\cite{LinPNAS2014,LiuPRL2016}. 
The estimate $\theta_{2D} \approx 0.52$ \cite{LiuPRL2016} is displayed as a dash line in
the inset of Fig.~\ref{fig:PXvaryingDamping}, which also illustrates the size dependence
of $\Px$ affecting the cutoff at small values of $x$.

When lowering the dissipation rate $\Gamma$ a systematic change in the  behavior
of  $\Px$ is observed, with the development of a steeper gap in $\Px$ at small values of $x$.
Basically, as inertial avalanches start to dominate, the probability of surviving
(not yielding) with a small value of $x$ (i.e., very close to the local threshold)
decreases.
Small barriers will be overcome during a massive avalanche. 
An apparent tendency to preserve a behavior of the form $\Px\sim x^\theta$
at low $x$ is observed, with an exponent $\theta$ that would increase as damping lowers.
This seems to contrast with the behavior of $\PS$ and $\PT$ that show
a clear characteristic peak appearing as inertia becomes relevant
(making possible a distinction between two kinds of avalanches).
However, the situation is a bit more complex.
In fact, the steeper growth at the smallest values of $x$ in each curve
is indeed a trace of the presence of two
kind of events.
The fact that $\Px$ considers the full set of local values of a configuration
in contrast to $S$ or $T$ that only provide a global avalanche property,
just renders more difficult the possibility of visualizing the avalanche
heterogeneity at this point.

If one insists on thinking the largest power-law growth for each
curve Fig.\ref{fig:PXvaryingDamping}, as a damping-modified
marginal stability scenario where only the exponent $\theta$
changes, depending on $\Gamma$, we can find no simple explanation 
for that exponent $\theta(\Gamma)$.
In fact, due to the emergence of the upper peak in $\PS$, the scaling
relation linking $\tau$, $\df$ and $\theta$ derived in~\cite{LinPNAS2014} 
for the overdamped case,
i.e. $\tau=2-\frac{\theta}{\theta+1}\frac{d}{\df}$,
is no longer expected to be valid in the inertial case.
Instead, we show below that the behavior displayed by $\Px$
is the combined result of an alternation between events roughly
classified in two different kinds, those similar to the well-known
overdamped avalanches and those dominated by inertial effects.

\subsection{Identifying and splitting two types of avalanches}

\iffigures
\begin{figure}[t!]
\begin{center}
\includegraphics[width=0.9\columnwidth, clip]{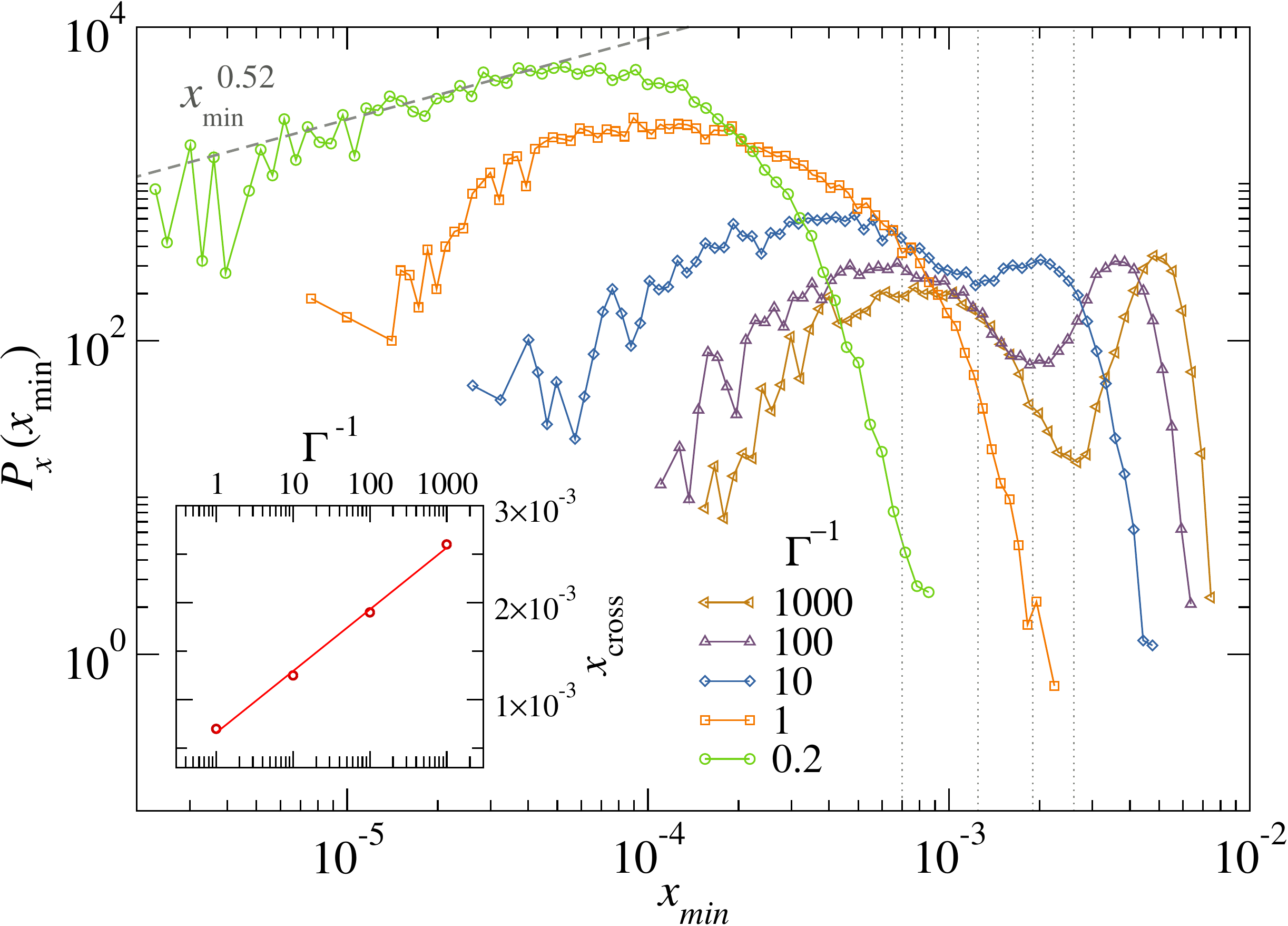}
\end{center}
\caption{\textit{Minimal distance to yielding distributions varying damping}.
$\Px(\xmin)$ versus $\xmin$ at different inverse dissipation coefficients $\Gamma^{-1}$.
Data correspond to a linear system size $L=40$ (2046 blocks).
In the overdamped limit $\Px(\xmin)$ copies the scaling of $\Px(x)$,
as is expected from independent sampling arguments.
As inertia becomes important $\Px(\xmin)$ acquires a bimodal shape, clearly separating
events that have left the system far away from instability from those that are similar
to the ones observed in the overdamped case.
Dashed gray lines mark the particular value of $\xmin$ at which we locate this crossover, $\xcross$
Inset: $\xcross$ as a function of $\Gamma^{-1}$ represented by open circles.
The full line is a fitting with a logarithmic function, yielding:
$\xcross \simeq 0.00066 + 0.000276 \ln(\Gamma^{-1})$.
} 
\label{fig:PXminVaryingDamping}
\end{figure}
\fi

In order to better understand the response of $\PS$, $\PT$ and $\Px$ to changes
in the damping coefficient, we analyze $\xmin \equiv \min \{ x_i \}$,
the quantity that determines the loading needed to trigger the next stress drop
after an avalanche ends.

Figure \ref{fig:PXminVaryingDamping} shows the distributions of such $\xmin$ values
during the stationary plastic flow for systems with different damping $\Gamma$.
In the overdamped limit we find that $\Px(\xmin)$ seems to initially
grow as $\xmin^\theta$ with $\theta \approx 0.52$, the same exponent
as for the initial grow of the full distribution $\Px(x)$
\footnote{For the minimal element $\xmin$ of independent samples $\{x\}$ taken from
$\Px(x)=x^\theta h(x)$ --with $h(x)$ a rapidly decaying function for $x\gg 1$--
a Weibull distribution is expected; whose initial grow at small values
of the argument follows a power-law with the same exponent $\theta$ left by the
original distribution.}.
This being anticipated by Karmakar \textit{et al.}~\cite{karmakar2010statistical},
was not seen instead in other works~\cite{LinEPL2014}, were $\Px(\xmin)$ seems
flat at small $\xmin$.
As inertia becomes important $\Px(\xmin)$ acquires a bimodal shape;
this is, two characteristic local maxima separated by an intermediate
local minimum.
Denoting by $\xcross$ the position of the minimum in $\Px(\xmin)$
(a saddle point for $\Gamma=1$), we find a logarithmic
dependence of $\xcross$ with $\Gamma^{-1}$.
$\xcross$ roughly separates two kinds of events:
On the one hand, those that have been presumably massive, affecting many sites, and have
left the system far away from instability, accumulating on a peak of ``large'' $\xmin$ values.
On the other hand, those events that we can consider to be similar to the ones in the overdamped case,
more localized avalanches, involving a small number of sites compared to the full system,
and leaving back some relatively weak spots without yielding.
A priori, we could link the characteristic ``large'' $\xmin$ values, with the typically big
stress drops observed in Fig.~\ref{fig:stressVSgamma}-left for the underdamped systems.

\iffigures
\begin{figure}[t!]
\begin{center}
  \begin{overpic}[width=0.9\columnwidth, clip]{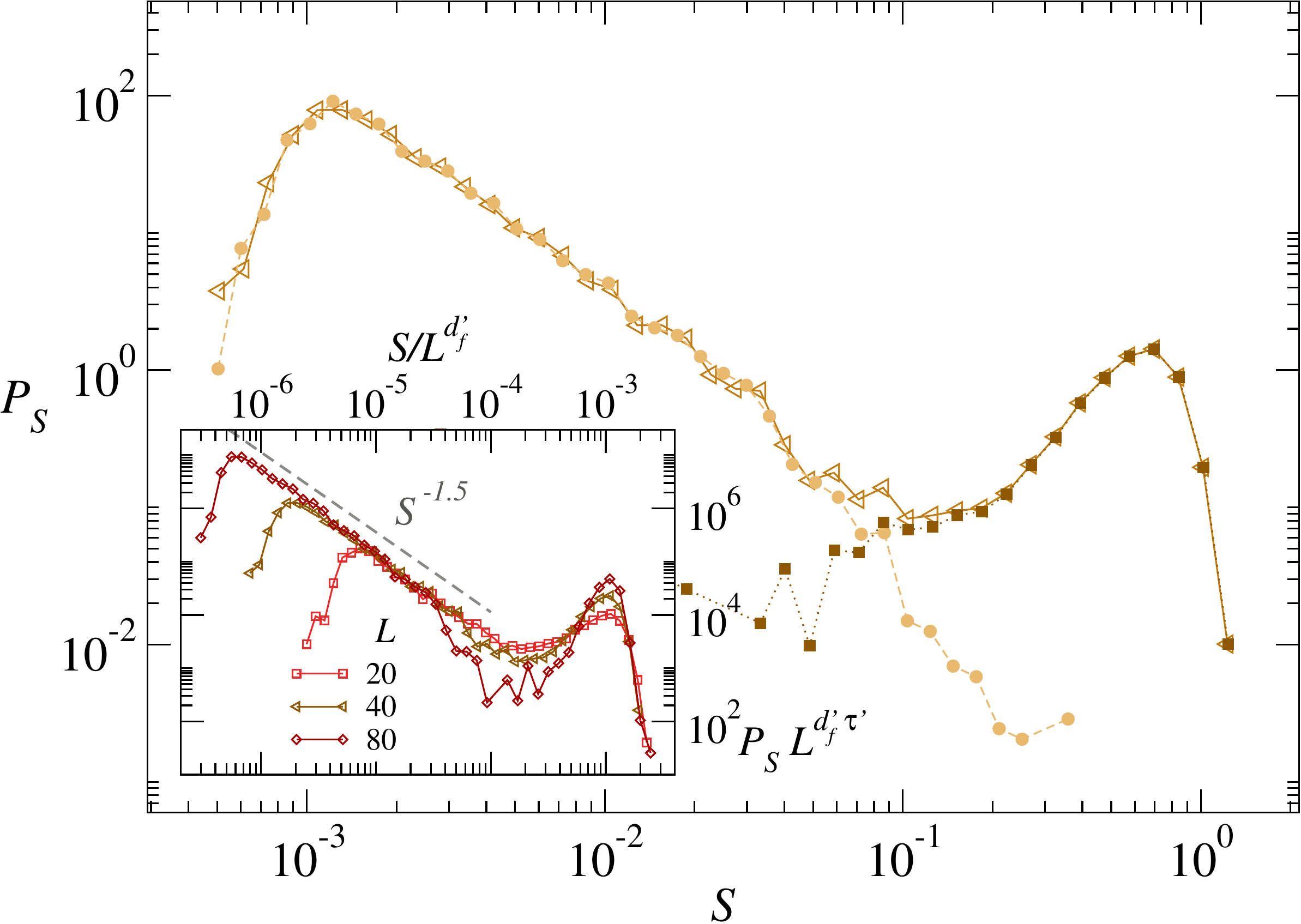}
     \put(75,47){\includegraphics[width=0.2\columnwidth]{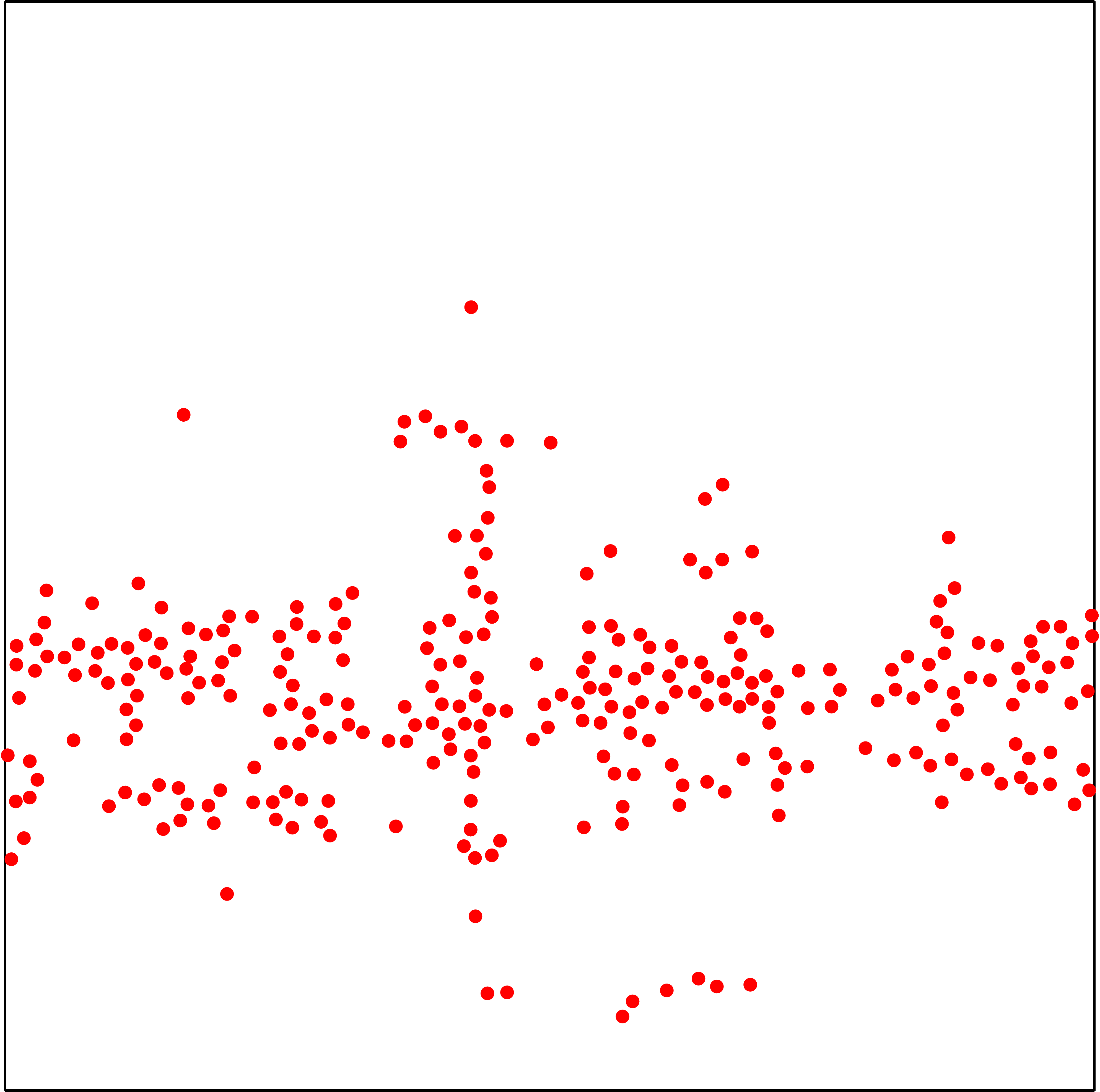}}  
     \put(51,47){\includegraphics[width=0.2\columnwidth]{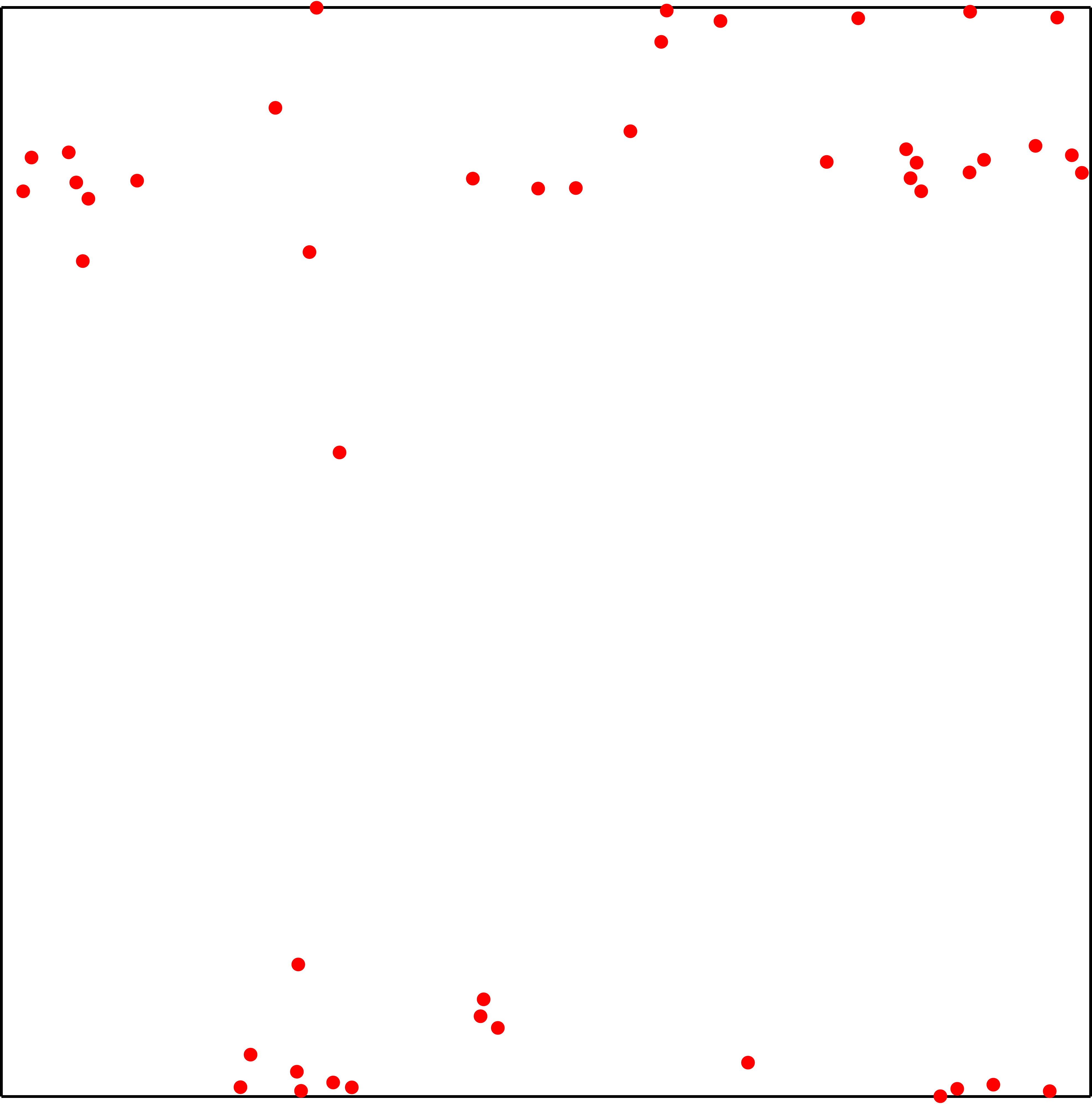}}  
  \end{overpic}
\end{center}
\caption{\textit{Avalanche size distributions split in two contributions}.
$\PS$ versus $S$ shown to be composed by two contributions, one coming from
the overdamped-like events and other one coming from the ``inertial'' events.
Data correspond to a linear system size $L=40$ (2046 blocks) and $\Gamma^{-1}=1000$.
Upper Insets:
Activity maps of selected avalanches.
Red dots represent points on the grid that were activated at least once
during the avalanche.
On the left, a typical avalanche corresponding to the overdamped sub-set.
On the right, an inertial avalanche.
Lower Inset:
$\PS$ shown for $\Gamma^{-1}=1000$ and different system sizes $L=20,40,80$.
An alignment for the position of the rightmost peak in $\PS$ is obtained
when rescaling $S$ by $L^{d'_f}$, with a damping dependent fractal dimension
$d'_f(\Gamma^{-1}=1000) \approx 1.75$.
} 
\label{fig:PofSwXminSplit}
\end{figure}
\fi

For the time being, let us take the proposed separation as an ansatz,
and test its validity with a self-consistency criterion.
We then use $\xcross$ to discriminate avalanches according to the $\xmin$ value
that they yield.

Figure~\ref{fig:PofSwXminSplit} shows in the main plot the $\PS$ distribution
for $L=40$ and the lowest damping simulated $\Gamma^{-1}=1000$.
The same curve as in Fig.\ref{fig:PSvaryingDamping} is shown by open triangles,
and has been built with the contribution of all avalanches in the set.
Now we discriminate avalanches yielding small values of $\xmin$ and avalanches yielding
large values of $\xmin$ and plot their distributions with filled light color
circles and filled dark color squares respectively.
After rescaling these contributions according to their weight in the full $\PS$, we
see that the discrimination between two kinds of avalanches, provided by $\xcross$
is quite accurate.
Other values of $\Gamma$ (not shown) display the same beautiful splitting of $\PS$
contributions, while, of course, the low-$\xmin$ contribution earns more and more
weight as  the overdamped limit is approached.

As it happens in studies of avalanches in spin systems~\cite{PerezRechePRB2003},
by accepting the idea of a separation between different kinds
of avalanches, one opens the door to go further and analyze
scaling properties of their respective distributions.
In the following we focus on some  scaling features of inertial avalanches
that contrast  with the overdamped case.

\subsection{Growing fractal dimension and incipient shear localization}
\label{sec:Growingfractaldimension}
%
Let us first briefly recall the finite-size scaling discussion of
Sec.\ref{sec:avalanche-distributions}.
The lower inset of Fig.\ref{fig:PofSwXminSplit} shows the same data as in 
Fig.~\ref{fig:PofSfinitesizeUnderdamped}, corresponding to $\Gamma^{-1}=1000$
and $L=20,40,80$, now rescaled as $\PS S^{d'_f \tau'}$ vs $S/L^{d'_f}$, where
we have used $d'_f=1.75$ and $\tau'=1.5$.
Notice that the inertial effect seems, in fact, stronger for bigger systems;
the relative peak height increases with $L$.
Of course, this does not happen for the equivalent scaling in the overdamped
limit (Fig.~\ref{fig:PSvaryingDamping}-inset), where the distribution cutoffs
identically collapse.
This tells us that the basic ingredient for the emergence of two different
kinds of collective events is, indeed, the lack of dissipation, and not the
finiteness of the box.

Now we turn our attention to the scaling chosen in the abscissas axis of
Fig.\ref{fig:PofSwXminSplit}-inset.
Here we have attempted to ``align'' the positions of the inertial peaks.
When doing so, we find a dependence $S_{\text{peak}}\sim L^{d'_f}$,
where $d'_f$ is a damping dependent fractal dimension for the avalanches
and governs the ``new'' cutoff of $\PS$.
In all cases $d'_f$ is bigger than the fractal dimension of the overdamped case.
An effective fractal dimension bigger than $d_f$ suggests that
the geometry should be very different between overdamped and inertial
avalanches.
Indeed, this is clearly manifested in the spatial coverage of a typical avalanche.
As insets of Fig.~\ref{fig:PofSwXminSplit} we show two images  representing
a system sample, where we have depicted activity maps of selected avalanches.
Red dots represent points on the grid that were activated at least once
during the avalanche.
On the left, a typical avalanche corresponding to the overdamped sub-set
shows a sparse quasi-1$d$ arrangement of active sites, consistent with a
fractal dimension $\df \approx 0.95$ as the one obtained from the finite
size scaling of overdamped systems~\cite{SalernoPRE2013,LiuPRL2016}.
On the right, an inertial avalanche belonging to the rightmost peak of $\PS$,
shows a much more dense and broad structure, compatible with a fractal dimension
closer to the dimension of the system ($d=2$).
When we inspect the largest avalanches for different damping coefficients,
we find that the generic form their pattern in space changes systematically,
becoming denser as $d'_f$ evolves from $0.95$ to $1.75$ in the studied regime.

Inertia, therefore, is found to modify the geometry and fractal dimension
of the avalanches at the yielding point, creating much denser events.
The geometry of these system spanning and broad avalanches are suggestive
of incipient shear bands.
In fact, very recent works both using finite elements methods~\cite{KarimiPRE2016} and
molecular dynamics~\cite{NicolasPRL2016}, have associated the absence of dissipation
with a non-monotonic flow-curve in the rheology of the system, and therefore mechanical
instability and the emergence of strain localization (see also~\cite{PapanikolaouPRE2016}).
Although we are working with a quasi-static protocol, the self-sustained effect
of inertial waves generating the inertial avalanches may be the same effect that generates
the shear bands in finite strain rate driving protocols.

\subsection{$\Px$ contributions and stability-geometry scaling relation}

\iffigures
\begin{figure}[t!]
\begin{center}
\includegraphics[width=0.9\columnwidth, clip]{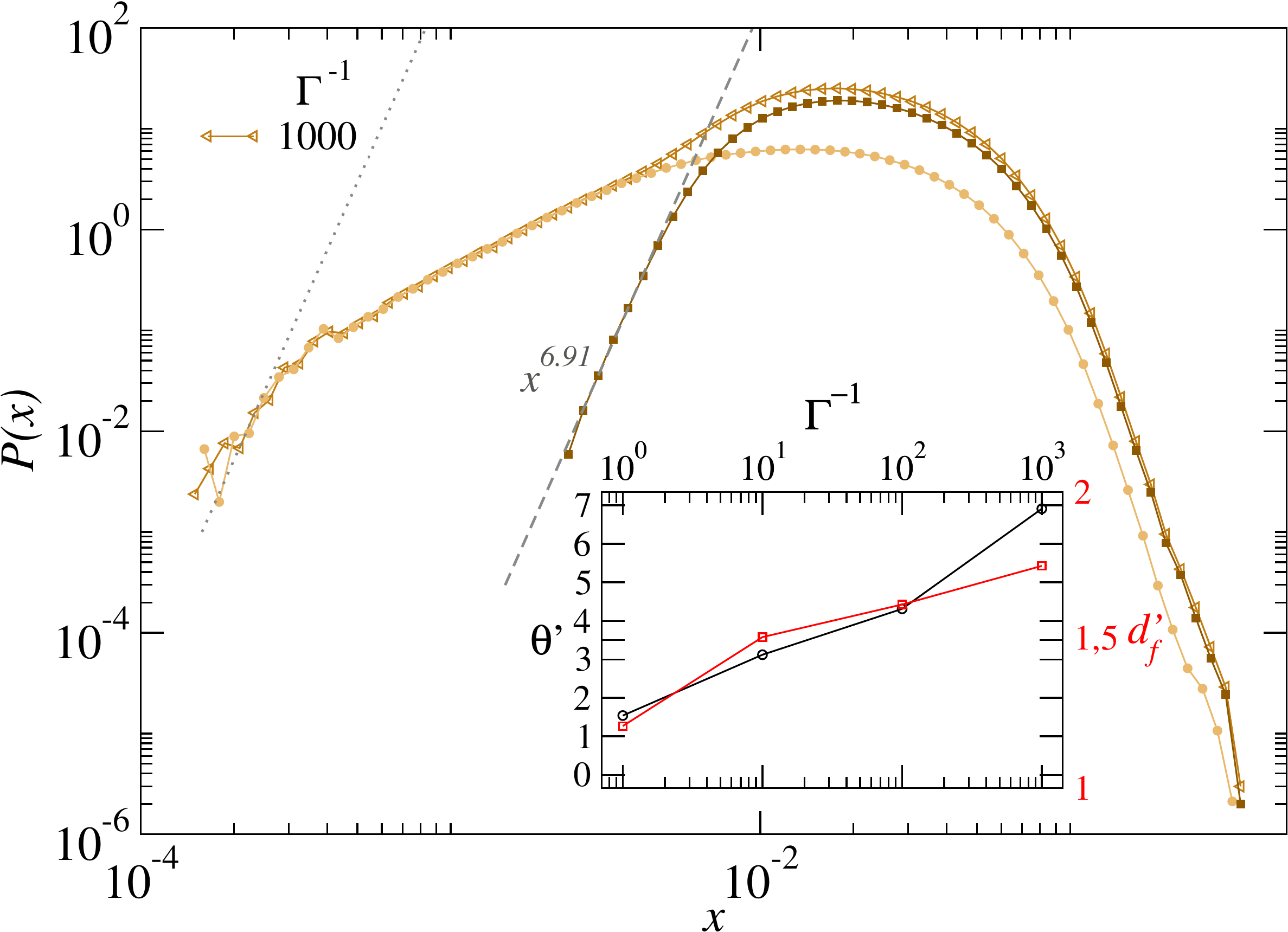}
\end{center}
\caption{\textit{Distances to yielding distributions varying damping split in two contributions}.
$\Px$ versus $x$ shown to be composed by two contributions, one coming from
the overdamped-like events and other one coming from the ``inertial'' events.
Data correspond to a linear system size $L=40$ (2046 blocks) and $\Gamma^{-1}=1000$
An alternative exponent $\theta'(\Gamma^{-1})$ can be obtained for the
law $\Px \sim x^{\theta'}$ when considering only inertial events.
The gray dashed line show this fit for $\Gamma^{-1}=1000$,
the dot dashed line replicates the same slope.
Inset:
$\theta'$ and the corresponding $d'_f = d \left( 1 - \frac{1}{1+\theta'} \right)$
as a function of $\Gamma^{-1}$.
} 
\label{fig:PofXwXminSplit}
\end{figure}
\fi

We now turn to the analysis of the different contributions to $\Px$.
Again, using the splitting criterion between
overdamped-like avalanches, that leave a $\xmin<\xcross$, and
inertial avalanches, that yield a $\xmin>\xcross$,
we plot the full-set distributions of distances to yielding $\Px$
together with two sub-set distributions.

Figure \ref{fig:PofXwXminSplit} show such a superposition of curves
for our lowest damping system ($\Gamma^{-1}=1000$).
We first notice that  the distribution $\Px$ for the sub-set of configurations
after an inertial avalanche has a much sharper  gap than the global 
distribution.
This is consistent, of course, with the fact that these events  typically have a larger
$\xmin$ value than the average $\xmin$.
In addition, a remarkable  finding is that 
the $\Px$ of inertial events display also a power-law growth at small $x$.
In other words, despite the massive events that push all local sites to have a
relatively large value of $x$ after the avalanche, these values are still  arranged in
a marginally stable fashion, as $\Px \sim x^{\theta'}$,
but now with a steeper exponent $\theta'> \theta_{\text{overdamped}}$.
In fact, assuming that the very general finite-size property
$\left<x_{\text{min}}\right>\sim L^{-\frac{d}{\theta'+1}}$~\cite{LinEPL2014,MullerARCMP2015}
also holds for this sub-set, and asking
$\left<x_{\text{min}}\right> \simeq \left<\Delta\sigma\right>$ during stationary plastic flow,
we expect
\begin{equation}\label{eq:scalingtheta}
 L^{-\frac{d}{\theta'+1}} \sim L^{d'_f - d} ~ \Rightarrow ~ \theta'= \frac{1}{1-(d'_f/d)}  - 1 
\end{equation}
where $d'_f$ is the exponent ruling the finite-size dependency of the peak position
in $\PS$ (see Fig.\ref{fig:PofSwXminSplit} inset), or equivalently, the scaling of
the sub-set of inertial avalanches.
 
The above relation between $\theta'$ and $d'_f$ holds well for all low-damping
systems studied in this work, as long as $\left<\Delta\sigma\right>$
is controlled by the inertial peak. This can be seen by  fitting $\theta'$ in the
inertial sub-set of the $\Px$ distribution.
For example, for $\Gamma^{-1}=1000$ we have $\theta'\simeq 6.91$ and $d'_f\simeq 1.75$,
for $\Gamma^{-1}=10$, $\theta'\simeq 3.12$ and $d'_f\simeq 1.5$.

Another interesting observation is that, even the subset of configurations
that are left behind by an overdamped-like avalanche show at the smaller
values of $x$  a growth that tends to be compatible with the same exponent
$\theta'$ (see dotted line, parallel to the dashed fit in Fig.\ref{fig:PofSwXminSplit}).
We interpret this feature as a fingerprint of inertial avalanches, with a depletion
($\theta' > \theta$)  in the amount of sites close to yielding even after one or 
several smaller avalanches has taken place.
This fingerprint of inertial avalanches could also explain the dependence on damping
of the exponent $\tau$ in the ``overdamped-like'' subsets of avalanche size distributions.
Even when short-duration avalanches do not directly \textit{feel} the effect of inertia,
they have to deploy correlations on a particularly heterogeneous landscape left behind by 
inertial avalanches.
As a result, the separation in two classes appears as a convenient classification,
but does not fully account for the complexity introduced by inertial effects.

\section{Summary}
\label{sec:summary}

We have analyzed the noise statistics of stress signals
produced by an amorphous solid under quasi-static deformation,
through numerical simulations of a realistic continuum model
treated with classical finite-element techniques.
In particular we have focused our analysis on the dependence
of such statistics with the ability of the system to dissipate energy,
spanning a wide range of damping values.

We validate our model by comparing our results for the distributions
of different observables in the overdamped limit with previously reported numerical
results in a variety of different models and techniques.

Both the distributions of avalanche sizes $\PS$ and durations $\PT$
display a growing peak at high values of $S$ and $T$, respectively,
when we lower the damping coefficient.
We have associated these characteristic peaks with a ``new'' kind of
avalanches, peculiar to inertial systems, which are triggered  and amplified by
elastic waves generated by other avalanches.
Such  events  are, in nature, more related
to effective thermal heating than to the deployment of large spatial
correlations, in line with~\cite{NicolasPRL2016}.
These  system spanning inertial avalanches are also characterized by a geometry
that is reminiscent of shear bands of strain
localization, although our protocol is quasi-static.

The distribution of minimal distances to yielding $\Px(\xmin)$ allowed
us to formulate  an \textit{ad hoc} criterion to discriminate
between these inertial avalanches and overdamped like events  that are much
more localized and not affected by the elastic wave propagation.
Using this criterion we have found an explanation for the
behavior at different damping of the full-set $\Px$ distribution,
that is considered as the steady state property that controls
the stability of the system and indirectly all the properties of 
its noise signal.
In particular, following very general arguments, we propose a 
scaling relation between the distribution $\Px$ of inertial avalanches
and their fractal dimension (Eq.~\ref{eq:scalingtheta}).

\section{Outlook}
\label{sec:discussion}

\subsection*{About the upper size cutoff}
As mentioned in the introduction, the irruption of inertial effects
in the --otherwise overdamped-- driven dynamics has been found to
break down the universal avalanche statistics of several systems.
It is worth stressing, though, an important difference between the
sand-pile problem with local yield thresholds and invariably positive
load redistributions (depinning models included), and the 
elasto-plastic models that describe plastic flow in solids:
Systems with a positive load redistribution produce avalanches that
are compact objects in space ($\df \geq d$).
The system size $L$ controls the cutoff for the avalanche sizes.
An overdamped system will explore this upper boundary displaying some
system-size spanning avalanches, but there is no possibility to observe bigger
avalanches than $S_c \sim L^\df$. 
When the overdamped condition is released in such systems, the avalanche size
distribution is modified~\cite{PradoPRA1992}.
The original power-law of the overdamped case deforms into a shorter scale-free
regime followed by a kink or peak that depends on the value of the damping.
However, for all damping choices, the largest avalanche accessible for a
given system size remains the same.

In the deformation of amorphous solids, instead, we have avalanches with $\df<d$
due to the heterogeneous (Eshelby) redistribution of stresses.
The cutoff of the avalanche distribution in overdamped systems is far from being
a massive avalanche involving all sites of the system.
In fact, both for 2D and 3D systems, $\df$ is found to be close to one~\cite{SalernoPRE2013,LiuPRL2016};
meaning that even a system-spanning avalanche leaves most of the sites untouched. 
Therefore, when inertia comes into play, the system still has \textit{room} to make
avalanches grow further.
This can be seen in Fig.~\ref{fig:PSvaryingDamping}.
As we lower the damping starting in the overdamped limit, first a plateau and
then a peak develops in $\PS$; all this happening to the \textit{right} of the
overdamped system-size cutoff.
May this explain why, for moderate damping, inertial effects are not strongly
evident in yield stress systems and avalanche distributions remain quite
similar to the one of the overdamped case, with occasionally the added features
of a small plateau at large sizes and a weak change in the $\tau$ exponent.
A picture that makes it appealing to conclude about a still universal scenario,
slightly modified by inertia~\cite{SalernoPRL2012,SalernoPRE2013}.

However, going deeper in the inertial regime while understanding the system size role,
we realize that indeed critical behavior breaks down while a damping-dependent typical
event size emerges in the form of a very clear peak in $\PS$.
Even though a scale-free power-law regime remains observable and is only weakly
affected by inertia, the inertial peak increasingly dominates the statistics of events.
Furthermore, since systems that dissipate energy at different rates show dissimilar
avalanche distributions, even when characterized by some scaling exponents,
we find it inaccurate to talk about universality.

\subsection*{Connections to other non universal statistics}
We have seen that the characteristic avalanche distribution of underdamped systems
in our prescription is a superposition of two populations.
Inertial and overdamped-like events interleave in time producing a unique statistics
that we cannot discriminate beforehand.
Both kinds of events are results of the same dynamical rules and boundary conditions.
It simply happens that from the competition of different time scales present in
the dynamics, both kinds emerge.
Even more, sometimes we cannot tell if one event is of one kind or the other.

Such a situation is not unique to the introduction of inertial effects.
It can also be observed in cases where a second time scale is introduced
by viscoelasticity~\cite{LandesPRL2014} or retardation~\cite{PapanikolaouPRE2016}.
Indeed, a quasi-periodic oscillation of the stress field was
argued in~\cite{LandesPRL2014} to be a possible explanation for deviations
from a pure power law (Gutenberg-Richter (GR) like) in the distribution
of earthquake magnitudes.
In fact, some years ago, a  discussion arose in the seismology community
contrasting opposite models of earthquake statistics:
On one hand the famous GR power-law decay kind of distribution;
on the other hand, the ``characteristic earthquake'' hypothesis predicting
a time recurrence of typically big earthquakes of a characteristic magnitude
on each individual fault (see~\cite{ben2008collective,deArcangelisPR2016} and references therein).
In spite of a marked predilection for a pure GR picture, the discussion remains
somewhat open~\cite{parsons2009there,page2015southern,deArcangelisPR2016},
and specific models with such characteristics are developed for
single faults~\cite{BarbotScience2012}.
We do not pretend to accredit here our simple model with relevance to the phenomenology
of earthquakes, but to highlight the ubiquity of the ``characteristic event'' feature.
In line with previous works~\cite{CarlsonPRL1989,CarlsonPRA1989,CarlsonPRA1991},
our model shows how a single set of dynamical rules can lead to the
emergence of distinct kind of events, ones with no characteristic size
and a Gutenberg-Richter distribution, and others with a magnitude that fluctuates
around a typical value in a peaked fashion.
For example, the role played by inertia in our model can be compared
to the idea of ``dynamic triggering'' of earthquakes~\cite{JohnsonNature2005},
which has attracted considerable attention in the last ten years.
In the present case we see that dynamical amplification through sound
waves, rather than triggering, appears as a dominant mechanism.
Our large avalanches could however be understood as consisting of an
initial event dynamically triggering a series of aftershocks at remote distances
through the wave propagation, the total stress drop magnitude being
the outcome of the whole series.
 
Finally, we note that laboratory systems such as metallic glasses also display statistics that
deviate from a pure power law with exponential cutoff, as can be seen from the inspection of
cumulative distributions shown for example in reference~\cite{AntonagliaPRL2014}.
Making similar studies in granular suspensions, in which inertial effects can be controlled
by using solvents of various viscosities, would be of great interest.
Interestingly, recent experiments on granular layers sheared between elastic
plates~\cite{Geller2015}, intended as a model for an isolated strike-slip
earthquake fault, also indicate a separation between two classes of events,
with large, system spanning events emerging in the tail of a broad
continuum power-law spectrum.

\subsection*{Concluding remarks}
Overall, inertia and amplification through sound waves  appears as a possible mechanism for
enriching the statistics of intermittent phenomena in deformed solids, with deviations from
universal power law statistics and large system spanning events that should be connected with
the possibility of shear band formation in these systems.
We hope that this work may stimulate experimental studies of intermittent behavior in systems
that may display strong inertial effects and/or strain localization, and we also consider
generalizing this study to finite strain rates in the future.

\begin{acknowledgments}
The authors acknowledge financial support from ERC grant ADG20110209.
JLB is supported by IUF.
Most of the computations were performed using the Froggy platform of the 
\href{https://ciment.ujf-grenoble.fr}{CIMENT} infrastructure supported by the Rh\^one-Alpes region (GRANT CPER07-13
\href{http://www.ci-ra.org/}{CIRA}) and the Equip@Meso project (reference ANR-10-EQPX-29-01).
Further we would like to thank Alexandre Nicolas, Alberto Rosso and J\'er\^ome Weiss  for a careful reading and useful feedback.
\end{acknowledgments}

\bibliographystyle{apsrev4-1}
\bibliography{avalanche}


\end{document}